\documentstyle[amssymb,preprint,aps]{revtex}

\begin{document}
\draft
\title{Andreev reflection through a quantum dot\\
coupled with two ferromagnets and a superconductor}
\author{{\ Yu Zhu, Qing-feng Sun and }Tsung-han Lin$^{*}$}
\address{{\it State Key Laboratory for Mesoscopic Physics and }\\
{\it Department of Physics, Peking University,}{\small \ }{\it Beijing}\\
100871, China}
\date{}
\maketitle

\begin{abstract}
We study the Andreev reflection (AR) in a three terminal mesoscopic hybrid
system, in which two ferromagnets (F$_1$ and F$_2$) are coupled to a
superconductor (S) through a quantum dot (QD). By using non-equilibrium
Green function, we derive a general current formula which allows arbitrary
spin polarizations, magnetization orientations and bias voltages in F$_1$
and F$_2$. The formula is applied to study both zero bias conductance and
finite bias current. The current conducted by crossed AR\ involving F$_1$, F$%
_2$ and S is particularly unusual, in which an electron with spin $\sigma $
incident from one of the ferromagnets picks up another electron with spin $%
\bar{\sigma}$ from the other one, both enter S and form a Cooper pair.
Several special cases are investigated to reveal the properties of AR in
this system.
\end{abstract}



\newpage

\baselineskip 20pt 

\section{INTRODUCTION}

Electron has spin as well as charge. The application of the electron spin
property opens a fruitful field in the transport of ferromagnetic materials,
such as the discovery of GMR and TMR effects \cite{MR}. On the other hand,
there are growing interests on the mesoscopic normal-metal / superconductor
(N/S) hybrid system \cite{hybrid}, in which Andreev reflection (AR) at N/S
interface plays an important role in low bias voltage regime \cite{btk}. In
AR process, an electron incident with energy E and spin $\sigma $ picks up
another electron with energy -E and spin $\bar{\sigma}$, both enter S and
form a Cooper pair, leaving a Andreev reflected hole in N side. One may
expect that the interplay of the spin property of AR process and the
spin-dependent transport in ferromagnetic materials will add new physics to
mesoscopic hybrid systems, and to the future applications of spintronics.

Several works have been devoted to this issue. In the pioneering work of de
Jong ${\sl et}$ ${\sl al}$. \cite{fs1}, the transport of a ferromagnet /
superconductor (F/S) junction was studied by scattering matrix formalism.
The conductance of AR is shown to be strongly affected by the spin
polarization of F. The idea was verified by recent experiments in F/S thin
film nanocontact \cite{fs2} and F/S metallic point contact \cite{fs3}.
Especially, in Ref. \cite{fs3}, Soulen ${\sl et}$ ${\sl al}$. successfully
determined the spin polarization at the Fermi energy for several metals by
measuring the differential conductance of F/S metallic point contact.
Further calculations \cite{fs4} implied that the Fermi velocity mismatch
between F and S also affects AR conductance of F/S contact, and the
conductance may even be enhanced in presence of spin polarization. In
addition to simple F/S junction, F/S contact with S in d-wave symmetry \cite
{fs4,fs(d)}, F/S nanostructure with giant proximity effect \cite{fspx1,fspx2}%
, and more complicated structures such as FSF double junctions \cite
{fsf1,fsf2,fsf3}, SFS\ double junctions \cite{sfs1,sfs2,sfs3}, S/F
superlattices \cite{sfsfs}, (NF)$_n$S multilayer structures \cite
{nfnfns1,nfnfns2} were also investigated.

In this paper, we propose an idea that two sources of spin polarized
electrons with different orientations are injected into a superconductor,
which can be achieved by a three terminal mesoscopic F/S hybrid structure
shown in Fig.1. In this system, a central quantum dot (QD) is coupled via
tunnel barriers to two ferromagnetic electrodes (F$_1$ and F$_2$) and a
superconducting electrode (S) ( hereafter, the system is simply referred as
to (F$_1$,F$_2$)-QD-S ). F$_1$ and F$_2$ are assumed to have arbitrary
magnetization orientations, spin polarizations, and bias voltages. The bias
voltage of S is set to zero as the ground. QD is designed to provide a link
between F$_1,$ F$_2$ and S, so that AR can take place through discrete
energy states of QD. Consider the special case that F$_1$ and F$_2$ are
fully polarized, AR\ only involving F$_1$ and S or only involving F$_2$ and
S are completely suppressed, while the crossed AR involving F$_1$, F$_2$,
and S depends strongly on the magnetization orientations of F$_1$ and F$_2$,
being suppressed if they are in ferromagnetic alignment, enhanced in
anti-ferromagnetic alignment. In this paper, we will derive a current
formula by using non-equilibrium Green function method, and investigate
several special cases to illustrate the properties of ARs in this system.

During the preparation of this paper, we are aware that in the recent
publication of Deutscher ${\sl et}$ ${\sl al}$. \cite{f2s}, a device
consisting of two point contacts between two ferromagnetic tips and a
superconductor was proposed. For the two tips with fully but opposite spin
polarizations, they suggested that ``mixed'' Cooper pair made of electrons
coming one from each tip can be injected into the superconductor, leading to
unusual properties of such device. Sec.IV of this paper is partially
stimulated by their work.

The rest of this paper is organized as follows: In Sec.II, we present the
model Hamiltonian and derive a general current formula for the hybrid system
(F$_{1,}$F$_2$)-QD-S, by non-equilibrium Green function method. In Sec.III,
we study the zero bias conductance, assuming $V_1=V_2=0^{+}$. The explicit
forms of the conductance are presented and numerically studied. In Sec.IV,
we study the finite bias current with F$_1$ and F$_2$ in anti-ferromagnetic
alignment, and the fully spin polarized case is discussed in detail.
Finally, a brief summary is given in Sec.V.

\section{MODEL AND FORMULATION}

The system under consideration can be described by the following Hamiltonian:

\begin{eqnarray}
H &=&H_1+H_2+H_{dot}+H_s+H_T\;\;, \\
H_1 &=&\sum_{k\sigma }(\epsilon _k-\sigma h_1-\mu _1)a_{k\sigma }^{\dagger
}a_{k\sigma }\;\;,  \nonumber \\
H_2 &=&\sum_{k\sigma ^{\prime }}(\epsilon _k-\sigma ^{\prime }h_2-\mu
_2)b_{k\sigma ^{\prime }}^{\dagger }b_{k\sigma ^{\prime }}\;\;,  \nonumber \\
H_{dot} &=&E_0\sum_\sigma c_\sigma ^{\dagger }c_\sigma \;\;,  \nonumber \\
H_s &=&\sum_{p\sigma }\epsilon _pd_{p\sigma }^{\dagger }d_{p\sigma
}+\sum_p\left[ \Delta d_{p\uparrow }^{\dagger }d_{-p\downarrow }^{\dagger
}+h.c\right] \;\;,  \nonumber \\
H_T &=&\sum_{k\sigma }\left[ t_{1\sigma }a_{k\sigma }^{\dagger }c_\sigma
+h.c\right] +\sum_{k\sigma }\left[ t_{2\sigma }b_{k\sigma }^{\dagger
}c_\sigma +h.c\right] +\sum_{p\sigma }\left[ t_sd_{p\sigma }^{\dagger
}c_\sigma +h.c\right] \;\;.  \nonumber
\end{eqnarray}
$H_1$ and $H_2$ are the Hamiltonians of F$_1$ and F$_2$ in the mean field
approximation, with different magnetization orientations and chemical
potentials. The spin bands of F$_1$ (F$_2$) are split by $2h_1$ ($2h_2$) due
to the exchange energy. The magnetization orientation of F$_1$ is set as z
axis, while the orientation of F$_2$ as z$^{\prime }$ axis which has an
angle $\theta $ with respect to z axis. The operators with the
spin-quantization axis z and the operators with the spin-quantization axis z$%
^{\prime }$ are related by D$^{\frac 12}$ matrix as 
\begin{equation}
{b_{k\uparrow ^{\prime }}^{\dagger } \choose b_{k\downarrow ^{\prime }}^{\dagger }}%
=\left( 
\begin{array}{cc}
\cos \frac \theta 2 & -\sin \frac \theta 2 \\ 
\sin \frac \theta 2 & \cos \frac \theta 2
\end{array}
\right) 
{b_{k\uparrow }^{\dagger } \choose b_{k\downarrow }^{\dagger }}%
\;\;.
\end{equation}
$H_{dot}$ describes the quantum dot, in which only one spin degenerate level
is considered and the intra-dot interaction is ignored for simplicity. $H_s$
is the Hamiltonian for a BCS\ superconductor with the chemical potential
fixed to zero as the ground. $H_T$ depicts the tunneling between QD and F$_1$%
, F$_2$ and S, coupling different parts of the system together.

Since the current through QD can be expressed in terms of the Green
functions of QD, we first derive the retarded and distribution Green
functions by Dyson equation and Keldysh equation. To include the physics of
Andreev reflections and the spin flip processes in a unified formulation, we
introduce a 4$\times $4 matrix representation, in which the Green function
is defined as 
\begin{equation}
{\bf G\equiv }\langle \langle \left( 
\begin{array}{l}
c_{\uparrow }^{\dagger } \\ 
c_{\downarrow } \\ 
c_{\downarrow }^{\dagger } \\ 
c_{\uparrow }
\end{array}
\right) \;|\;\left( 
\begin{array}{llll}
c_{\uparrow } & c_{\downarrow }^{\dagger } & c_{\downarrow } & c_{\uparrow
}^{\dagger }
\end{array}
\right) \rangle \rangle \;\;.
\end{equation}

Let ${\bf G}^r$ denote the Fourier transformed retarded Green function of QD
, and ${\bf G}^r$ can be solved by Dyson equation: 
\begin{equation}
{\bf G}^r{\bf =g}^r{\bf +g}^r{\bf \Sigma }^r{\bf G}^r\;\;,
\end{equation}
in which ${\bf g}^r$ is the retarded Green function of an isolated QD and $%
{\bf \Sigma }^r$ is the self-energy due to couplings between QD and leads. $%
{\bf g}^r$ can be easily obtained as: 
\begin{equation}
{\bf g}^r=\left( 
\begin{array}{cccc}
\frac 1{\omega -E_0+\text{i}0^{+}} & 0 & 0 & 0 \\ 
0 & \frac 1{\omega +E_0+\text{i}0^{+}} & 0 & 0 \\ 
0 & 0 & \frac 1{\omega -E_0+\text{i}0^{+}} & 0 \\ 
0 & 0 & 0 & \frac 1{\omega +E_0+\text{i}0^{+}}
\end{array}
\right) \;\;,
\end{equation}
while ${\bf \Sigma }^r$ consists of three parts, 
\begin{equation}
{\bf \Sigma }^r={\bf \Sigma }_1^r+{\bf \Sigma }_2^r+{\bf \Sigma }_s^r\;\;.
\end{equation}
${\bf \Sigma }_1^r$ is the self-energy from the coupling between QD and F$_1$%
, given by 
\begin{equation}
{\bf \Sigma }_1^r=-\frac{\text{i}}2\left( 
\begin{array}{cccc}
\Gamma _{1\uparrow } & 0 & 0 & 0 \\ 
0 & \Gamma _{1\downarrow } & 0 & 0 \\ 
0 & 0 & \Gamma _{1\downarrow } & 0 \\ 
0 & 0 & 0 & \Gamma _{1\uparrow }
\end{array}
\right) \;\;,
\end{equation}
in which $\Gamma _{1\uparrow }$ and $\Gamma _{1\downarrow }$ are the
spin-dependent coupling strengths defined by $\Gamma _{1\sigma }\equiv 2\pi
N_{1\sigma }\left| t_{1\sigma }\right| ^2$, with $N_{1\sigma }$ being the
density of states of spin $\sigma $ band of F$_{1.}$ ${\bf \Sigma }_2^r$ is
the self-energy from the coupling between QD and F$_2$, given by 
\begin{equation}
{\bf \Sigma }_2^r=-\frac{\text{i}}2\left( 
\begin{array}{cccc}
c^2\Gamma _{2\uparrow }+s^2\Gamma _{2\downarrow } & 0 & sc(\Gamma
_{2\uparrow }-\Gamma _{2\downarrow }) & 0 \\ 
0 & c^2\Gamma _{2\downarrow }+s^2\Gamma _{2\uparrow } & 0 & sc(\Gamma
_{2\uparrow }-\Gamma _{2\downarrow }) \\ 
sc(\Gamma _{2\uparrow }-\Gamma _{2\downarrow }) & 0 & c^2\Gamma
_{2\downarrow }+s^2\Gamma _{2\uparrow } & 0 \\ 
0 & sc(\Gamma _{2\uparrow }-\Gamma _{2\downarrow }) & 0 & c^2\Gamma
_{2\uparrow }+s^2\Gamma _{2\downarrow }
\end{array}
\right) \;\;,
\end{equation}
in which $s\equiv $ $\sin \frac \theta 2$ , $c\equiv \cos \frac \theta 2$ , $%
\Gamma _{2\uparrow }$ and $\Gamma _{2\downarrow }$ are defined similarly to $%
\Gamma _{1\uparrow }$ and $\Gamma _{1\downarrow }$. ${\bf \Sigma }_s^r$ is
the self-energy from by the coupling between QD and S, given by 
\begin{equation}
{\bf \Sigma }_s^r=-\frac{\text{i}}2\Gamma _s\rho (\omega )\left( 
\begin{array}{cccc}
1 & -\frac \Delta \omega & 0 & 0 \\ 
-\frac \Delta \omega & 1 & 0 & 0 \\ 
0 & 0 & 1 & \frac \Delta \omega \\ 
0 & 0 & \frac \Delta \omega & 1
\end{array}
\right) \;\;,
\end{equation}
in which $\Gamma _s\equiv 2\pi N_s\left| t_s\right| ^2$, with $N_s$ being
the density of states when the superconductor is in normal state, $\rho
(\omega )$ is the modified BCS density of states defined by 
\begin{equation}
\rho (\omega )\equiv \left\{ 
\begin{array}{cc}
\frac{|\omega |}{\sqrt{\omega ^2-\Delta ^2}} & \;\;\;\;|\omega |>\Delta \\ 
\frac \omega {\text{i}\sqrt{\Delta ^2-\omega ^2}} & \;\;\;\;|\omega |<\Delta
\end{array}
\right. \;\;.
\end{equation}
Thus, ${\bf G}^r$ can be obtained by solving Dyson equation, Eq.(4).

Let ${\bf G}^{<}$ denote the Fourier transformed distribution Green function
of QD, and ${\bf G}^{<}$ can be obtained by Keldysh equation: 
\begin{equation}
{\bf G}^{<}{\bf =G}^r{\bf \Sigma }^{<}{\bf G}^a\;\;.
\end{equation}
Notice that the advanced Green function and self-energy are the Hermitian
conjugations of the corresponding retarded Green function and self-energy.
And ${\bf \Sigma }^{<}$ can be obtained by applying the
fluctuation-dissipation theorem to each of ${\bf \Sigma }_1^{<}$, ${\bf %
\Sigma }_2^{<}$ and ${\bf \Sigma }_s^{<}$, 
\begin{eqnarray}
{\bf \Sigma }^{<} &=&{\bf \Sigma }_1^{<}+{\bf \Sigma }_2^{<}+{\bf \Sigma }%
_s^{<}\;\;, \\
{\bf \Sigma }_1^{<} &=&{\bf F}_1({\bf \Sigma }_1^a-{\bf \Sigma }_1^r)\;\;, 
\nonumber \\
{\bf \Sigma }_2^{<} &=&{\bf F}_2({\bf \Sigma }_2^a-{\bf \Sigma }_2^r)\;\;, 
\nonumber \\
{\bf \Sigma }_s^{<} &=&{\bf F}_s({\bf \Sigma }_s^a-{\bf \Sigma }_s^r)\;\;, 
\nonumber
\end{eqnarray}
in which 
\begin{equation}
{\bf F}_1=\left( 
\begin{array}{cccc}
f_1 & 0 & 0 & 0 \\ 
0 & \bar{f}_1 & 0 & 0 \\ 
0 & 0 & f_1 & 0 \\ 
0 & 0 & 0 & \bar{f}_1
\end{array}
\right) \;\;,
\end{equation}
\begin{equation}
{\bf F}_2=\left( 
\begin{array}{cccc}
f_2 & 0 & 0 & 0 \\ 
0 & \bar{f}_2 & 0 & 0 \\ 
0 & 0 & f_2 & 0 \\ 
0 & 0 & 0 & \bar{f}_2
\end{array}
\right) \;\;,
\end{equation}
\begin{equation}
{\bf F}_s=\left( 
\begin{array}{cccc}
f & 0 & 0 & 0 \\ 
0 & f & 0 & 0 \\ 
0 & 0 & f & 0 \\ 
0 & 0 & 0 & f
\end{array}
\right) \;\;,
\end{equation}
where $f_1$, $\bar{f}_1$, $f_2$, $\bar{f}_2$ and $f$ denote $f(\omega -V_1)$%
, $f(\omega +V_1)$, $f(\omega -V_2)$, $f(\omega +V_2)$ and $f(\omega )$,
respectively, in which $f(\omega )$ is the Fermi distribution function.

Then, the current flowing from F$_1$ to the QD can be expressed in terms of $%
{\bf G}^r$ and ${\bf G}^{<}$ as 
\begin{equation}
I_1=I_{1\uparrow }+I_{1\downarrow }=\frac eh\int d\omega \left[ \left( {\bf %
G\Sigma }_1\right) ^{<}+h.c\right] _{11+33}\;\;,
\end{equation}
in which we have used the compact notations $\left[ {\bf AB}\right] ^{<}{\bf %
\equiv A}^r{\bf B}^{<}{\bf +A}^{<}{\bf B}^a$ and $\left[ \;\;\right]
_{11+33}\equiv \left[ \;\;\right] _{11}+\left[ \;\;\right] _{33}$. After
some algebra manipulations, the current can be divided into contributions
from four conducting processes: 
\begin{equation}
I_1=\frac eh\int d\omega \left[ A_{11}(f_1-\bar{f}_1)+A_{12}(f_1-\bar{f}%
_2)+Q_{1s}(f_1-f_s)+Q_{12}(f_1-f_2)\right] \;\;,
\end{equation}
in which 
\begin{equation}
A_{11}=\Gamma _{1\downarrow }(\Gamma _{1\uparrow }\left| G_{12}^r\right|
^2+\Gamma _{1\downarrow }\left| G_{32}^r\right| ^2)+\Gamma _{1\uparrow
}(\Gamma _{1\uparrow }\left| G_{14}^r\right| ^2+\Gamma _{1\downarrow }\left|
G_{34}^r\right| ^2)
\end{equation}
represents the Andreev reflection through F$_1$-QD-S, 
\begin{eqnarray}
A_{12} &=&(c^2\Gamma _{2\downarrow }+s^2\Gamma _{2\uparrow })(\Gamma
_{1\uparrow }\left| G_{12}^r\right| ^2+\Gamma _{1\downarrow }\left|
G_{32}^r\right| ^2)+ \\
&&(c^2\Gamma _{2\uparrow }+s^2\Gamma _{2\downarrow })(\Gamma _{1\uparrow
}\left| G_{14}^r\right| ^2+\Gamma _{1\downarrow }\left| G_{34}^r\right| ^2)+
\nonumber \\
&&sc(\Gamma _{2\uparrow }-\Gamma _{2\downarrow })2%
\mathop{\rm Re}%
(\Gamma _{1\uparrow }G_{12}^rG_{14}^{r*}+\Gamma _{1\downarrow
}G_{32}^rG_{34}^{r*})  \nonumber
\end{eqnarray}
represents the crossed Andreev reflection through (F$_1$,F$_2)$-QD-S, 
\begin{eqnarray}
Q_{1s} &=&\Gamma _{1\uparrow }\Gamma _s\tilde{\rho}\left[ \left|
G_{11}^r\right| ^2+\left| G_{12}^r\right| ^2+\left| G_{13}^r\right|
^2+\left| G_{14}^r\right| ^2+2%
\mathop{\rm Re}%
(-\frac \Delta \omega G_{11}^rG_{12}^{r*}+\frac \Delta \omega
G_{13}^rG_{14}^{r*})\right] + \\
&&\Gamma _{1\downarrow }\Gamma _s\tilde{\rho}\left[ \left| G_{31}^r\right|
^2+\left| G_{32}^r\right| ^2+\left| G_{33}^r\right| ^2+\left|
G_{34}^r\right| ^2+2%
\mathop{\rm Re}%
(-\frac \Delta \omega G_{31}^rG_{32}^{r*}+\frac \Delta \omega
G_{33}^rG_{34}^{r*})\right]   \nonumber
\end{eqnarray}
represents the single particle tunneling through F$_1$-QD-S, and $\tilde{\rho%
}(\omega )\equiv \frac{\left| \omega \right| }{\sqrt{\omega ^2-\Delta ^2}}%
\theta (\left| \omega \right| -\Delta )$ is the ordinary BCS density of
states, 
\begin{eqnarray}
Q_{12} &=&(c^2\Gamma _{2\uparrow }+s^2\Gamma _{2\downarrow })(\Gamma
_{1\uparrow }\left| G_{11}^r\right| ^2+\Gamma _{1\downarrow }\left|
G_{31}^r\right| ^2)+ \\
&&(c^2\Gamma _{2\downarrow }+s^2\Gamma _{2\uparrow })(\Gamma _{1\uparrow
}\left| G_{13}^r\right| ^2+\Gamma _{1\downarrow }\left| G_{33}^r\right| ^2)+
\nonumber \\
&&sc(\Gamma _{2\uparrow }-\Gamma _{2\downarrow })2%
\mathop{\rm Re}%
(\Gamma _{1\uparrow }G_{11}^rG_{13}^{r*}+\Gamma _{1\downarrow
}G_{33}^rG_{31}^{r*})  \nonumber
\end{eqnarray}
represents the single-particle tunneling through F$_1$-QD-F$_2$. Similarly,
one can obtain the current flowing from F$_2$ into QD simply by exchange the
index 1 and 2.

The current formula Eq.(17) is the central result of this work, which can be
applied to ferromagnetic electrodes F$_1$ and F$_2$ with arbitrary spin
polarizations, magnetization orientations and bias voltages.

In the following numerical studies, we assume that $\left| eV_1\right|
,\left| eV_2\right| <\Delta $ and $k_BT\ll \Delta $. $Q_{1s}$ process will
vanish because of the factor $\tilde{\rho}$ and the Fermi function
difference $(f_1-f_s)$. $Q_{12}$ process will be ruled out in two special
cases: F$_1$ and F$_2$ are either equally biased (Section III) or fully but
oppositely polarized (Section IV). We will concentrate on AR processes $%
A_{11}$ (direct AR through F$_1$-QD-S) and $A_{12}$ (crossed AR through (F$%
_1 $,F$_2$)-QD-S ), and investigate several special cases to illustrate the
properties of these two ARs.

\section{ZERO BIAS CONDUCTANCE}

In this section, we study the zero bias conductance by taking $V_1=V_2=0^{+}$%
. Since no bias voltage between F$_1$ and F$_2$, there is no net
single-particle current flowing between them. For $k_BT\ll \Delta $, the
single-particle current from F$_1$ or F$_2$ to S is also negligible.
Therefore, only ARs contribute to the conductance. For simplicity, we set $%
k_BT=0$ and $E_0=0$, introduce the spin polarization $P_\beta \equiv \frac{%
\Gamma _{\beta \uparrow }-\Gamma _{\beta \downarrow }}{\Gamma _{\beta
\uparrow }+\Gamma _{\beta \downarrow }}$, and the spin-averaged coupling
strength $\Gamma _\beta \equiv \frac 12(\Gamma _{\beta \uparrow }+\Gamma
_{\beta \downarrow })$, with $\beta =$ $1,2$ for F$_1$and F$_2$ respectively.

First consider the simplest case in which $\Gamma _2=0$, $\Gamma _1\equiv
\Gamma _L$, $P_1\equiv P$, $\Gamma _s\equiv \Gamma _R$, then the
three-terminal system (F$_1,$F$_2$)-QD-S reduces to a two-terminal system
F-QD-S, and the conductance is easily obtained from the current formula as 
\begin{equation}
G_{FDS}=\frac{4e^2}h\frac{(1-P^2)r^2}{(1-P^2+r^2)^2}\;\;,
\end{equation}
where $r\equiv \Gamma _R/\Gamma _L$ is the ratio of the two coupling
strengths. Analogous to the matching condition of the Fermi velocities in
F/S contact (i.e., $k_{F\uparrow }k_{F\downarrow }=k_S^2),$ here $P^2+r^2=1$
(i.e., $\Gamma _{L\uparrow }\Gamma _{L_{\downarrow }}=\Gamma _R^2$) plays
the similar role. For $r>1$, the matching condition can never be satisfied,
so $G_{FDS}$ decreases monotonously with the increase of $P$ (Fig.2a)$.$
While for $r<1$ , it exists a certain value of $P$, say $P_0$, satisfying $%
P_0^2+r^2=1$, so $G_{FDS}$ first increases with $P$ , reaches its maximum
value $\frac{4e^2}h$ at $P=P_0$, then drops to 0 when $P$ approaches to 1
(Fig.2b). This result warns us to be careful to deduce the spin polarization
of F from AR conductance of F-QD-S.

Next, consider the general case of the three terminal system (F$_1,$F$_2$%
)-QD-S. Similar to the composition of polarized light, the total current (or
total conductance) of F$_1$ and F$_2$ are equivalent to that of an effective
ferromagnet \~{F}. Introduce the spin polarization vectors {\bf $\vec{q}$}$%
_1 $ and {\bf $\vec{q}$}$_2$, where {\bf $\vec{q}$}$_\beta $ has the
magnitude of $\Gamma _\beta P_\beta $ and the direction of the magnetization
direction of F$_\beta $, with $\beta =1,2$. It is easy to test that these
vectors obey the vector composition rule, i.e., {\bf $\vec{q}$}$=${\bf $\vec{%
q}$}$_1+${\bf $\vec{q}$}$_2$, in which {\bf $\vec{q}$ }is the spin
polarization vector of \~{F}. Therefore, the effective parameters of \~{F}
are 
\begin{eqnarray}
\tilde{\Gamma} &=&\Gamma _1+\Gamma _2\;\;, \\
\tilde{P} &=&\frac{\left[ (\Gamma _1P_1)^2+(\Gamma _2P_2)^2+2\Gamma
_1P_1\Gamma _2P_2\cos \theta \right] ^{\frac 12}}{\Gamma _1+\Gamma _2}\;\;. 
\nonumber
\end{eqnarray}
As a result, the total conductance of F$_1$ and F$_2$ can be obtained as 
\begin{equation}
G\equiv G_1+G_2=G_{FDS}(\tilde{P},\tilde{r})\;,
\end{equation}
in which $G_{FDS}$ has the same form as in Eq.(22), $\tilde{P}$ is the
effective polarization, and $\tilde{r}$ is defined by $\Gamma _s/\tilde{%
\Gamma}$. Then the conductance of F$_1$ and F$_2$ can be expressed by the
total conductance multiplied by a sharing factor, 
\begin{eqnarray}
G_1 &=&G\frac{\Gamma _1^2+\Gamma _1\Gamma _2-(\Gamma _1^2P_1^2+\Gamma
_1P_1\Gamma _2P_2\cos \theta )}{\Gamma _1^2+\Gamma _2^2+2\Gamma _1\Gamma
_2-(\Gamma _1^2P_1^2+\Gamma _2^2P_2^2+2\Gamma _1P_1\Gamma _2P_2\cos \theta )}%
\;\;, \\
G_2 &=&G\frac{\Gamma _2^2+\Gamma _1\Gamma _2-(\Gamma _2^2P_2^2+\Gamma
_1P_1\Gamma _2P_2\cos \theta )}{\Gamma _1^2+\Gamma _2^2+2\Gamma _1\Gamma
_2-(\Gamma _1^2P_1^2+\Gamma _2^2P_2^2+2\Gamma _1P_1\Gamma _2P_2\cos \theta )}%
\;\;.
\end{eqnarray}

Fig.3 shows the curves of $G$ vs $\theta $ (also can be viewed as $2G_1$ vs $%
\theta $ or $2G_2$ vs $\theta $ ) for the symmetric case, in which $\Gamma
_1=\Gamma _2\equiv \Gamma $ and $P_1=P_2\equiv P$. For $r=1$, $G$ increases
with the increase of $\theta $ or decrease of $P$. For $r>1$, the curves of $%
G$ vs $\theta $ is qualitatively the same as those of $r=1$, but the
conductance is lowered and more sensitive to $P.$ For $r<1$, the variation
is more complicated: if $P^2<1-r^2$, $G$ decreases with the increase of $%
\theta $ or decrease of $P$; if $P^2>1-r^2$, $G$ has the maximum $\frac{4e^2}%
h$ at $\theta $ satisfying $(P\cos \frac \theta 2)^2=1-r^2$. These results
are readily understood by the new matching condition $\tilde{P}^2+r^2=1$
with the effective spin polarization $\tilde{P}=P\cos \frac \theta 2$.

Two points are noteworthy in the above result: (1) If F$_1$ and F$_2$ are
regarded as a whole, the effective polarization can be tuned continuously by
changing the angle of the mutual orientations, which is impossible for one
chosen ferromagnet. (2) For $r\geqslant 1$, the total conductance for the
two ferromagnets ins anti-ferromagnetic alignment is larger than that in
ferromagnetic alignment, which is completely different from the effect of
GMR or TMR. To describe this new effect of magneto-resistance, define the
ratio of Andreev reflected magnetic resistance ($ARMR$) in (F$_1,$F$_2$%
)-QD-S by 
\begin{equation}
ARMR\equiv \frac{G_{AF}-G_F}{G_{AF}+G_F}\;\;.
\end{equation}
and the curves of $ARMR$ vs $P$ for various $r$ are shown in Fig.4.

Fig.5 shows the curves of $G_1$ vs $\theta $ for an asymmetric case, in
which $P_1=1$, $P_2$ is arbitrary, and $\Gamma _1=\Gamma _2=\Gamma _s/2$.
Since F$_1$ is fully polarized, the conductance of F$_1$ is sensitive to the
spin polarization and orientation of F$_2$. For $P_2=0$, $G_1$ does not
depend on $\theta $; while for $P_2=1$, $G_1$ strongly depends on $\theta $,
with $G_1=0$ at $\theta =0$ and $G_1=\frac{4e^2}h$ at $\theta =\pi $. We
suggest that this effect can be applied to measure the spin polarization of F%
$_2$. In practice,one may chose a half-metal material as F$_1$, the
ferromagnetic material to be measured as F$_2$, and changing the spin
orientation of F$_1$ by applying an external magnetic field, then the spin
polarization of F$_2$ can be deduced from the weak / strong dependence of $%
G_1$ on $\theta $.

\section{FINITE BIAS CURRENT}

Now we turn to investigate the non-equilibrium transport of (F$_1,$F$_2$%
)-QD-S. For simplicity, we only consider the antiparallel orientation of F$%
_1 $ and F$_2$ (i.e., $\theta =\pi $), with finite but small bias voltages
(i.e., $\left| eV_1\right| <\Delta $ and $\left| eV_2\right| <\Delta $).
Notice that the self-energy becomes to block-diagonal due to $\theta =\pi $,
and the expression of current $I_1$ can be simplified as, 
\begin{eqnarray}
I_1 &=&\frac eh\int d\omega \left[ A_{11}(f_1-\bar{f}_1)+A_{12}(f_1-\bar{f}%
_2)+Q_{1s}(f_1-f_s)+Q_{12}(f_1-f_2)\right] \;\;, \\
A_{11} &=&\Gamma _{1\downarrow }\Gamma _{1\uparrow }\left| G_{12}^r\right|
^2+\Gamma _{1\uparrow }\Gamma _{1\downarrow }\left| G_{34}^r\right| ^2\;\;, 
\nonumber \\
A_{12} &=&\Gamma _{2\uparrow }\Gamma _{1\uparrow }\left| G_{12}^r\right|
^2+\Gamma _{2\downarrow }\Gamma _{1\downarrow }\left| G_{34}^r\right| ^2\;\;,
\nonumber \\
Q_{1s} &=&\Gamma _{1\uparrow }\Gamma _s\tilde{\rho}\left[ \left|
G_{11}^r\right| ^2+\left| G_{12}^r\right| ^2+2%
\mathop{\rm Re}%
(-\frac \Delta \omega G_{11}^rG_{12}^{r*})\right] +  \nonumber \\
&&\Gamma _{1\downarrow }\Gamma _s\tilde{\rho}\left[ \left| G_{33}^r\right|
^2+\left| G_{34}^r\right| ^2+2%
\mathop{\rm Re}%
(+\frac \Delta \omega G_{33}^rG_{34}^{r*})\right] \;\;,  \nonumber \\
Q_{12} &=&\Gamma _{2\downarrow }\Gamma _{1\uparrow }\left| G_{11}^r\right|
^2+\Gamma _{2\uparrow }\Gamma _{1\downarrow }\left| G_{33}^r\right| ^2\;\;. 
\nonumber
\end{eqnarray}

At zero temperature and in the low bias regime, the current of $Q_{1s}$
process vanishes. Further assuming that both F$_1$ and F$_2$ are fully
polarized, both $Q_{12}$ and $A_{11}$ process are also forbidden. Only the
process of $A_{12}$, i.e., crossed AR involving F$_1,$F$_2$ and S
contributes to the current. $I_1$ and $I_2$ are derived as 
\begin{equation}
I\equiv I_1=I_2=\frac eh\int d\omega \Gamma _{2\uparrow }\Gamma _{1\uparrow
}\left| G_{12}^r\right| ^2(f_1-\bar{f}_2)\;\;.
\end{equation}
Notice that $I_1=I_2$ holds even if $\Gamma _1\neq \Gamma _2$ and $V_1\neq
V_2$, because $I_1$ is pure spin $\uparrow $ current and $I_2$ is pure spin $%
\downarrow $ current while $I_1+I_2$ is required to be non-spin-polarized
current by the superconductor. For simplicity, we further assume that $%
\Gamma _{1\uparrow }=\Gamma _{2\uparrow }\equiv \Gamma _L$ ($\Gamma
_{1\downarrow }=\Gamma _{2\downarrow }=0$ due to $P_1=P_2=1$) and $\Gamma
_s\equiv \Gamma _R$, then the system (F$_1,$F$_2$)-QD-S is similar to a
special N-QD-S one, in which the two spin bands of N have different chemical
potentials controlled by $V_1$ and $V_2$. Define the transmission
probability of crossed AR by $T_{AR}(\omega )\equiv \Gamma _L^2\left|
G_{12}^r\right| ^2$, the current can be expressed as 
\begin{equation}
I=\frac eh\int_{-V_2}^{V_1}T_{AR}(\omega )d\omega \;\;.
\end{equation}

Notice that $T_{AR}(\omega )$ is an even function of $\omega $, the above
current formula implies that the sign of $I_1$ or $I_2$ is not determined by 
$V_1$ or $V_2$ but by $\frac 12(V_1+V_2)$. This is quite unusual because it
contains the case that $V_1>0$ and $V_2<0$ but $I_1=I_2>0$ (This unusual
property was first addressed in \cite{f2s}). Generally, for a three-terminal
system, one may expect that current flows out of the terminal with highest
voltage and into the terminal with lowest voltage. But for the current
conducted by crossed AR, the sign of current in each ferromagnetic terminal
is linked to the averaged chemical potential of the two, because two
ferromagnets cooperate with each other in this process, with total energy
balanced. Fig.6 illustrates the conducting process corresponding to the case
of $I_1=I_2>0$ with $\mu _1>0$ and $\mu _2<0$ but $\frac 12(\mu _1+\mu _2)>0$%
.

We ignore the energy structure of QD in Fig.6 for simplicity, however, the
current $I\equiv I_1=I_2$ depends strongly on the transmission probability
of QD. In fact, $I$ is the integral of $T_{AR}(\omega )$ over the range of $%
(-V_2,V_1)$. Fig7 shows the surfaces of $I(V_1,V_2)$ and corresponding $%
T_{AR}(\omega )$ spectrum for three typical cases of $\Gamma _L$ and $\Gamma
_R$. In Fig.7a, $\Gamma _L\ll \Gamma _R$, the spin degenerate level of QD is
hybridized to two Andreev bound states due to coupling with S, while the
coupling with F$_1$ and F$_2$ provides the small broadening to these bound
states. $T_{AR}$ has two peaks with the maximum of unity at each of the
Andreev bound states. Correspondingly, the surface of $I(V_1,V_2)$ has five
steps: the highest step corresponds to $(-V_2,V_1)$ covering both of the
peaks; the second step (including two patches) corresponds to $(-V_2,V_1)$
covering one of the peaks; the third step (including three patches)
corresponds to $(-V_2,V_1)$ or $(V_1,-V_2)$ covering none of the peaks; the
fourth step (including two patches) corresponds to $(V_1,-V_2)$ covering one
of the peaks; and the lowest step corresponds to $(V_1,-V_2)$ covering both
of the peaks. In Fig.7b, $\Gamma _L=\Gamma _R$, the Andreev bound states are
sufficiently broadened so that the two peaks in $T_{AR}$ merge into one. The
one peak structure of $T_{AR}$ spectrum corresponds to three step pattern in 
$I(V_1,V_2)$ surface. In Fig.7c, $\Gamma _L\gg \Gamma _R$, the resonant
level of QD is significantly broadened, as a result, $T_{AR}$ is small and
flat with tails at $\omega =\pm \Delta $. The structureless $T_{AR}$
spectrum corresponds to a plain in $I(V_1,V_2)$ surface, proportional to $%
\frac 12(V_1+V_2)$. In short, $T_{AR}$ spectrum can be extracted from the
measurement of $I(V_1,V_2)$ surface.

\section{ CONCLUSIONS}

In this paper, we have investigated the Andreev reflection in a (F$_1,$F$_2$%
)-QD-S system. By using the non-equilibrium Green function, a general
current formula is derived, allowing arbitrary spin polarizations,
magnetization orientations and bias voltages in F$_1$ and F$_2$. The formula
is applied to several special cases, revealing some interesting properties
of this system: (1) Analogous to the Fermi velocity mismatch in F/S contact,
the zero bias conductance in F-QD-S reaches its maximum $\frac{4e^2}h$ if
matching condition $\Gamma _{L\uparrow }\Gamma _{L\downarrow }=\Gamma _R^2$
satisfied. (2) For total current (conductance) of (F$_1$,F$_2$)-QD-S with $%
V_1=V_2$, the two ferromagnets F$_1$ and F$_2$ are equivalent to an
effective ferromagnet \~{F}, and the effective polarization $\tilde{P}$ can
be tuned by the angle between the spin orientations of F$_1$ and F$_2$. (3)
There is a new effect of magneto-resistance in (F$_1$,F$_2$)-QD-S (named as $%
ARMR$), in which the conductance for F$_1$ and F$_2$ in anti-ferromagnetic
alignment is larger than that in ferromagnetic alignment. Base on this
effect, a possible way to measure the spin polarization of ferromagnetic
material is proposed. (4) The non-equilibrium transport of this system is
quite unusual. Especially, if F$_1$ and F$_2$ are fully but opposite
polarized, the signs of current through F$_1$ and F$_2$ is determined by $%
\frac 12(V_1+V_2)$ rather than $V_1$ or $V_2$. Furthermore, the surface of $%
I(V_1,V_2)$ depends strongly on the AR transmission probability, which can
be applied to extract the latter. Finally, we believe that the suggested (F$%
_1,$F$_2$)-QD-S system is accessible of the up-date nano-technology, and we
are eager to see relevant experiment on such appealing system.

\section*{ACKNOWLEDGMENTS}

This project was supported by NSFC under grant No.10074001. We would like to
thank Y. Lu and Y. F. Yang for useful discussions. One of the authors (T.-H.
Lin) would also like to thank the support from the Visiting Scholar
Foundation of State Key Laboratory for Mesoscopic Physics in Peking
University.

\smallskip $^{*}$ To whom correspondence should be addressed.


\newpage

\section*{Figure Captions}

\begin{itemize}
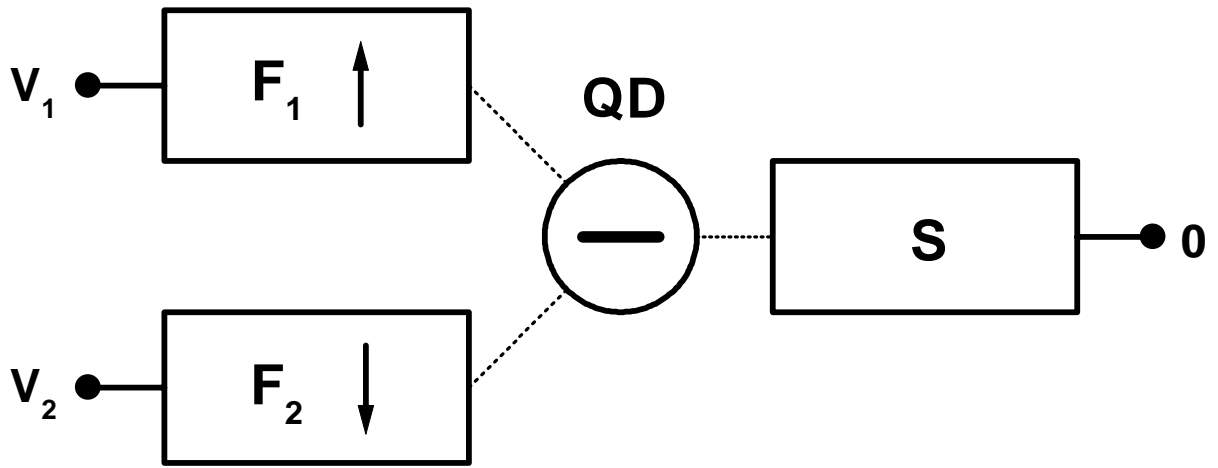
\item[{\bf Fig. 1}]  Schematic drawing of the three-terminal system under
consideration. F$_1$ and F$_2$ represent two ferromagnetic electrodes with
different magnetization orientations and bias voltages, QD is a quantum dot,
and S is a superconductor with zero voltage as the ground.

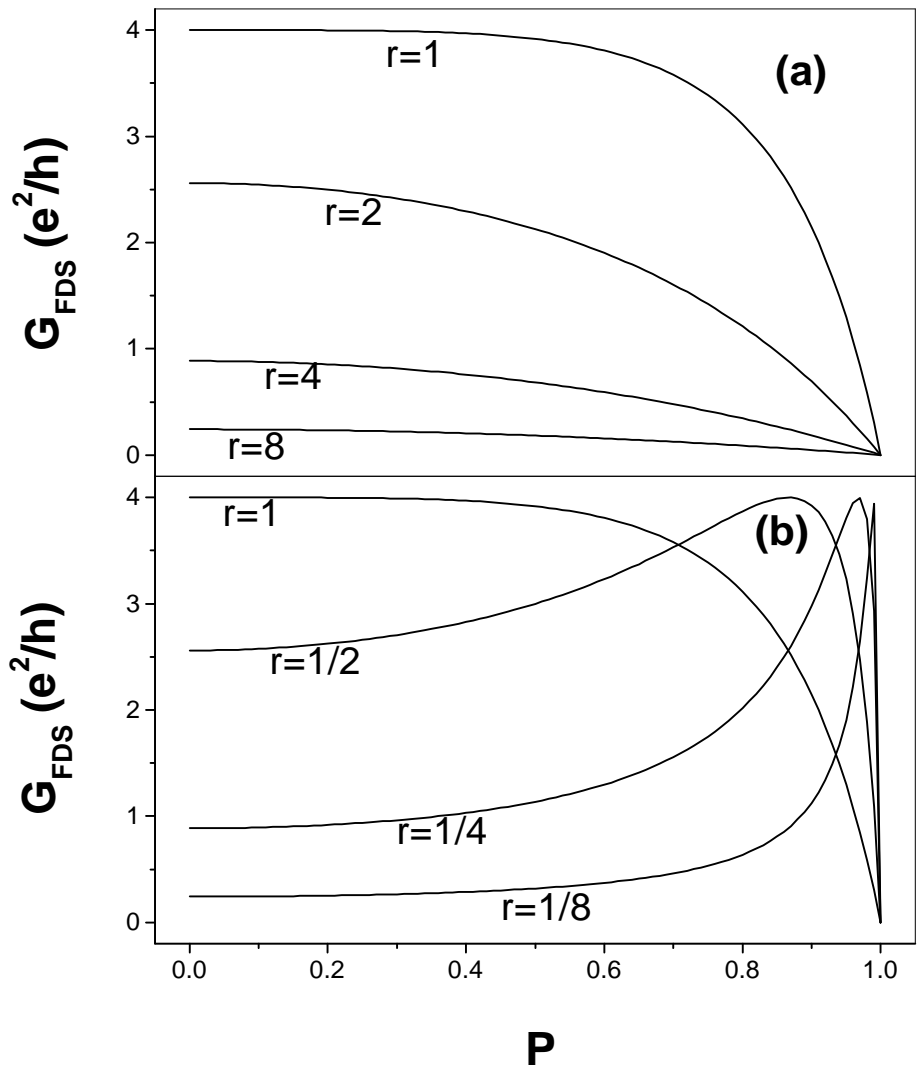
\item[{\bf Fig. 2}]  The zero bias conductance $G$ vs $P$ for F-QD-S, where $%
P$ is the spin polarization of F. $r\equiv \Gamma _R/\Gamma _L$ is the ratio
of coupling strengths, with $r\geqslant 1$ for (a) and $r\leqslant 1$ for
(b).

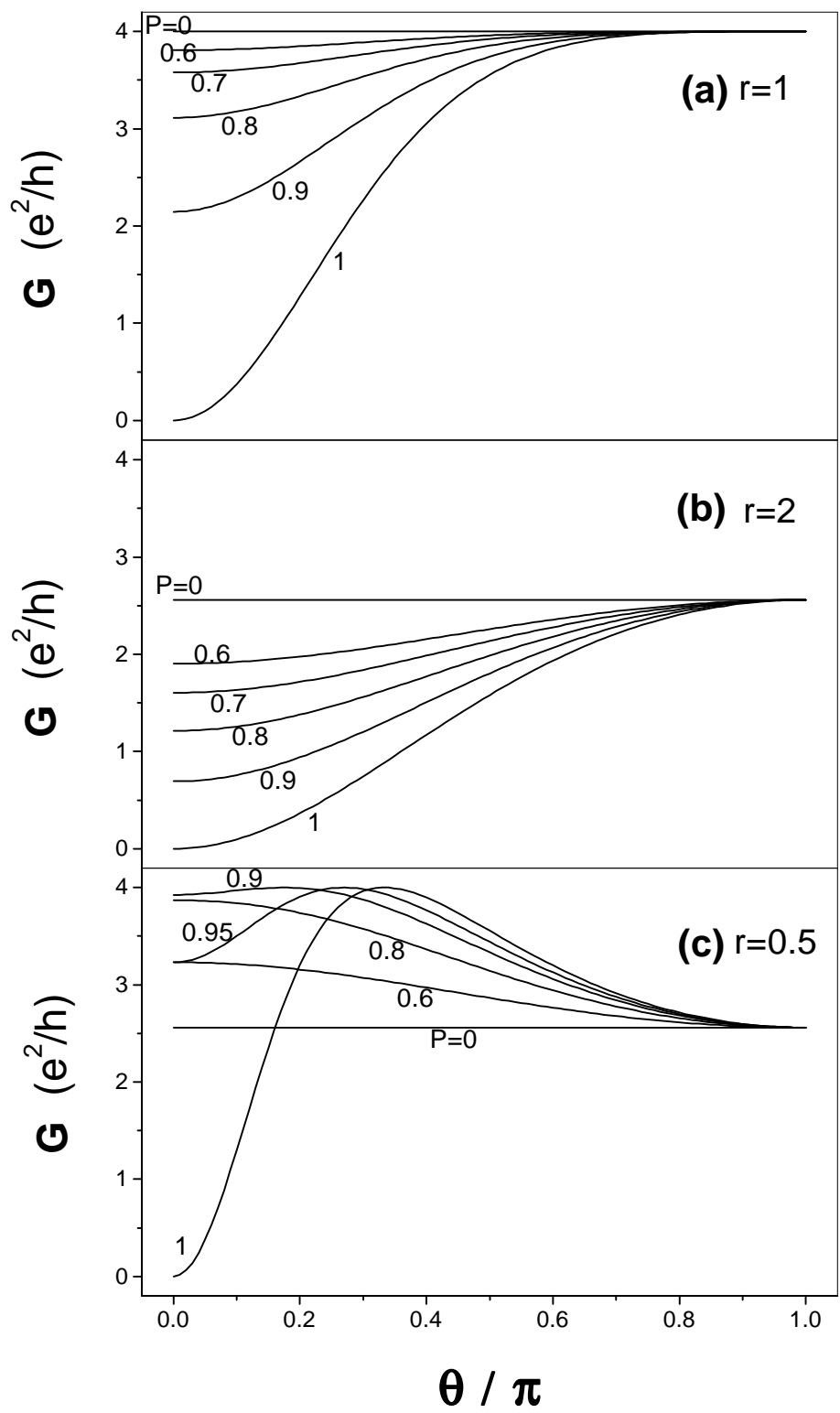
\item[{\bf Fig. 3}]  The total conductance $G$ vs $\theta $ for (F$_1,$F$_2$%
)-QD-S, where $\theta $ is the angle between the orientations of F$_1$ and F$%
_2$, with $\Gamma _1=\Gamma _2\equiv \Gamma $, $P_1=P_2\equiv P$, and $%
r\equiv \Gamma _0/(\Gamma _1+\Gamma _2)$.

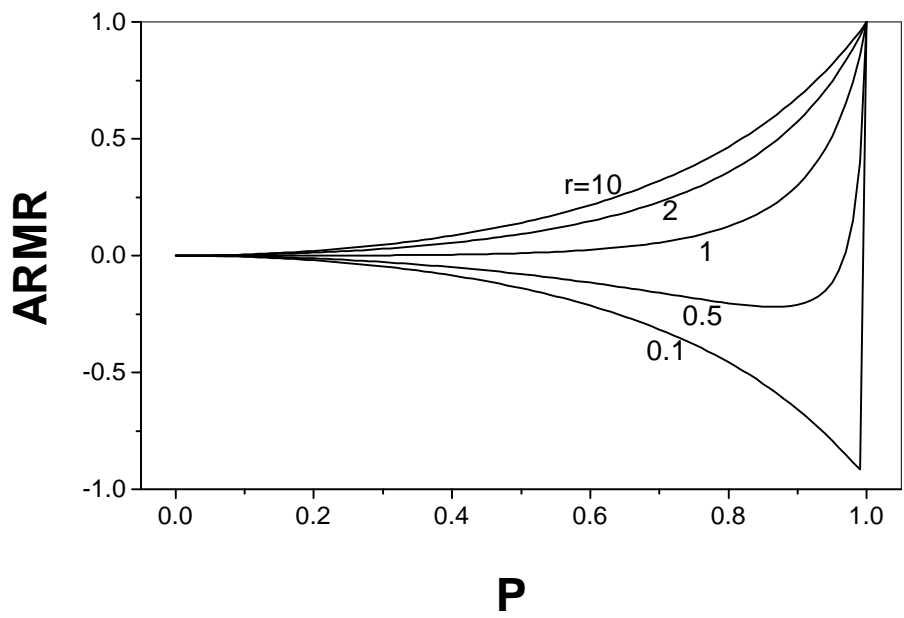
\item[{\bf Fig. 4}]  $ARMR$\ vs $P$ in (F$_1,$F$_2$)-QD-S, where $ARMR\equiv 
$\ $(G_{AF}-G_F)/(G_{AF}+G_F)$, $P$ and $r$ have the same meaning as in
Fig.3.

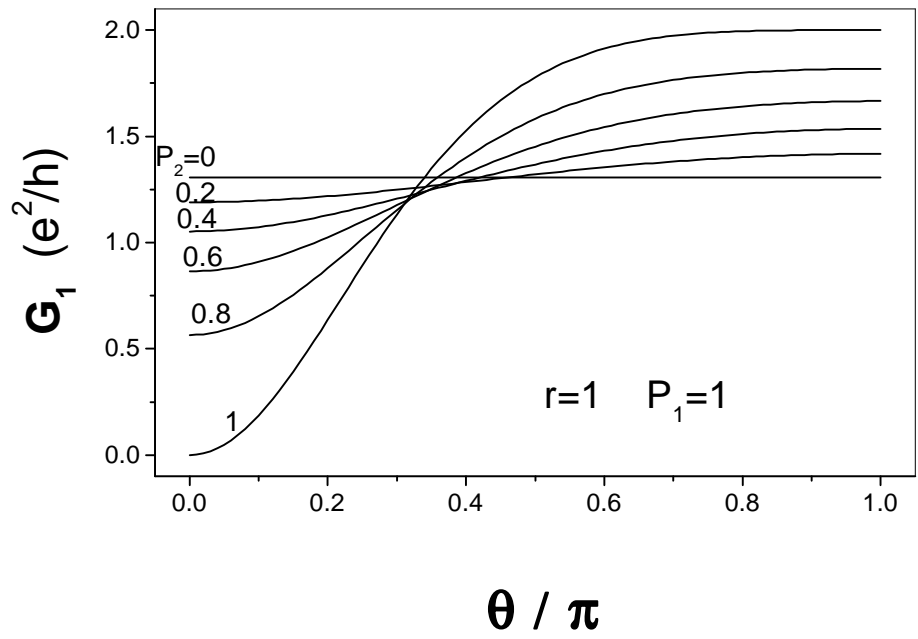
\item[{\bf Fig. 5}]  The conductance $G_1$ vs $\theta $ for different $P_2$,
with $P_1=1$ and $r=1$. $G_1$ has strong / weak dependence on $\theta $ for
large / small $P_2$, which can be applied to measure the spin polarization
of F$_2$. 

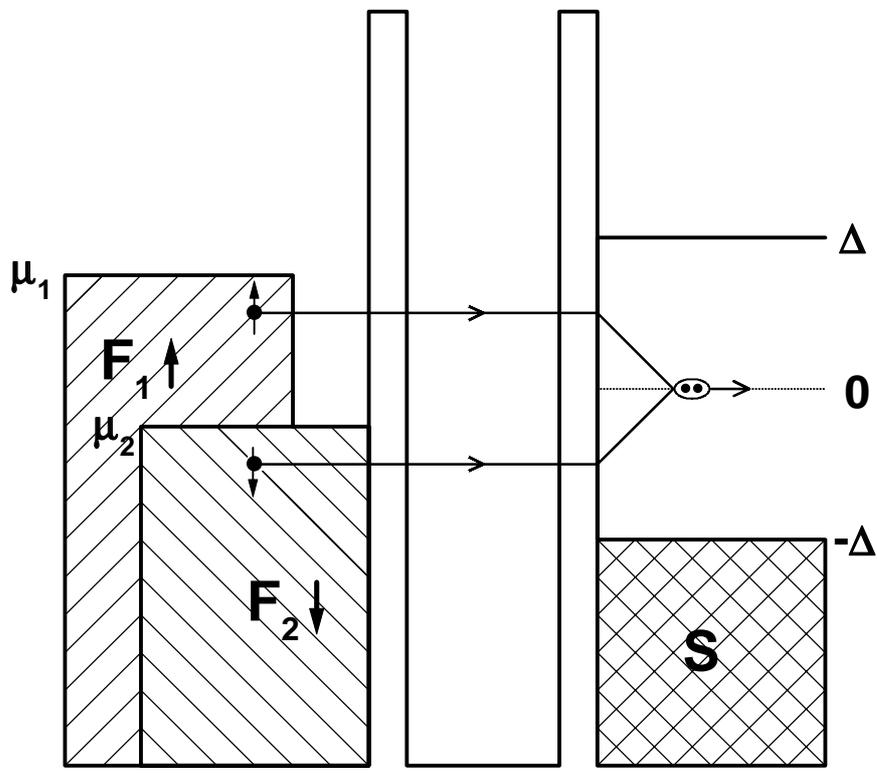
\item[{\bf Fig. 6}]  Schematic diagram of non-equilibrium transport in (F$_1,
$F$_2$)-QD-S. F$_1$ and F$_2$ are in anti-ferromagnetic alignment, marked by
left- and right- slanted shadows, respectively. S is marked by crossed
shadow, with the energy gap region $\pm \Delta $ with respect to the
chemical potential; QD is between the two barriers, and the energy structure
is ignored for simplicity. The diagram illustrates an unusual property of
the current conducted by crossed AR involving F$_1$, F$_2$ and S: the signs
of $I_1$ or $I_2$ are determined by $\frac 12(\mu _1+\mu _2)$ rather than $%
\mu _1$ or $\mu _2$.

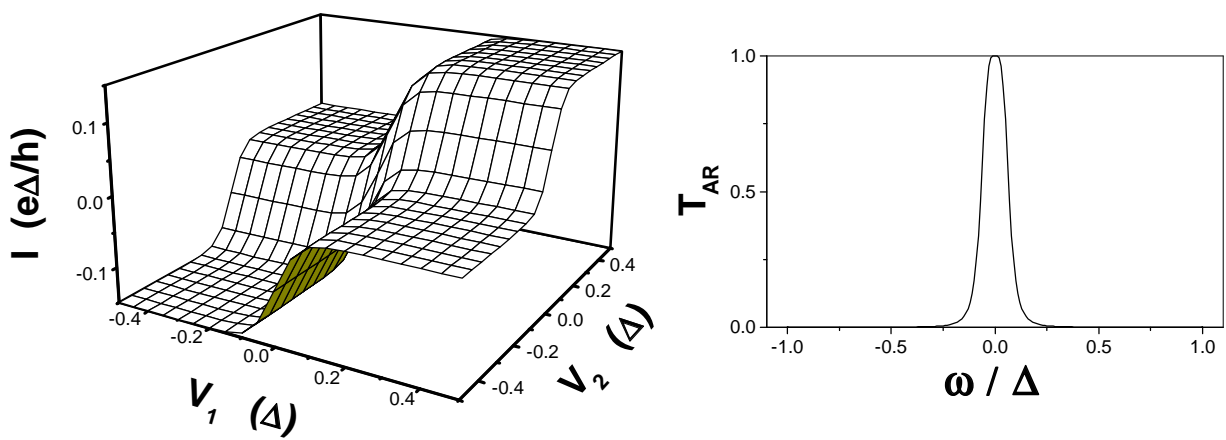
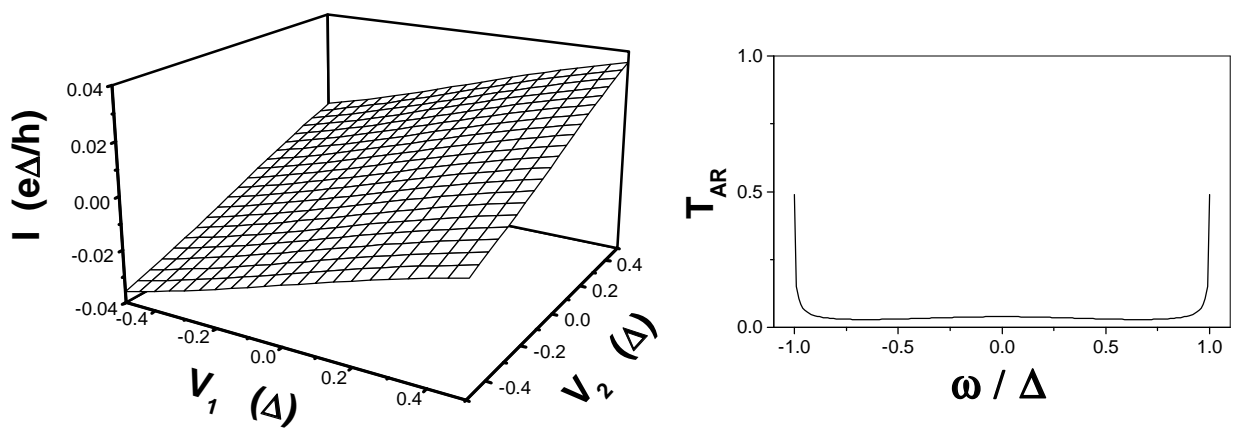
\item[{\bf Fig. 7}]  The current $I\equiv I_1=I_2$ vs the bias voltages $%
(V_1,V_2)$ for three typical cases: (a) $\Gamma _L\ll \Gamma _R$; (b) $%
\Gamma _L=\Gamma _R$; and (c) $\Gamma _L\gg \Gamma _R$. The surface of $%
I(V_1,V_2)$ has close relationship to the spectrum $T_{AR}(\omega )$, which
can be used to extract the latter.
\end{itemize}

\end{document}